\documentclass[twocolumn]{aastex61}
\usepackage{soul} 
\usepackage{graphicx}
\shorttitle{HIP 79977}
\shortauthors{Goebel et al.}

\begin{document}

\title{SCExAO/CHARIS Near-IR High-Contrast Imaging and Integral Field Spectroscopy of the HIP 79977 Debris Disk}

\correspondingauthor{Sean Goebel, Thayne Currie}
\email{sgoebel@ifa.hawaii.edu, currie@naoj.org}

\author{Sean Goebel}
\affiliation{Institute for Astronomy, University of Hawaii,
640 North A`oh$\bar{o}$k$\bar{u}$ Place, Hilo, HI 96720, USA}
\affiliation{Subaru Telescope, National Astronomical Observatory of Japan, 
650 North A`oh$\bar{o}$k$\bar{u}$ Place, Hilo, HI  96720, USA}
\author{Thayne Currie}
\affiliation{Subaru Telescope, National Astronomical Observatory of Japan, 
650 North A`oh$\bar{o}$k$\bar{u}$ Place, Hilo, HI  96720, USA}
\affiliation{NASA-Ames Research Center, Moffett Field, CA, USA}
\affiliation{Eureka Scientific, 2452 Delmer Street Suite 100, Oakland, CA, USA}
\author{Olivier Guyon}
\affiliation{Subaru Telescope, National Astronomical Observatory of Japan, 
650 North A`oh$\bar{o}$k$\bar{u}$ Place, Hilo, HI  96720, USA}
\affil{Steward Observatory, University of Arizona, Tucson, AZ 85721, USA}
\affil{College of Optical Sciences, University of Arizona, Tucson, AZ 85721, USA}
\affil{Astrobiology Center of NINS, 2-21-1, Osawa, Mitaka, Tokyo, 181-8588, Japan}
\author{Timothy D. Brandt}
\affiliation{Department of Physics, University of California, Santa Barbara, Santa Barbara, California, USA}
\author{Tyler D. Groff}
\affiliation{NASA-Goddard Space Flight Center, Greenbelt, MD, USA}
\author{Nemanja Jovanovic}
\affiliation{Caltech Optical Observatory, Department of Astronomy, California Institute of Technology, 1200 E. California Blvd., Pasadena, CA 91125, USA}
\author{N. Jeremy Kasdin}
\affiliation{Department of Mechanical Engineering, Princeton University, Princeton, NJ, USA}
\author{Julien Lozi}
\affiliation{Subaru Telescope, National Astronomical Observatory of Japan, 
650 North A`oh$\bar{o}$k$\bar{u}$ Place, Hilo, HI  96720, USA}
\author{Klaus Hodapp}
\affiliation{Institute for Astronomy, University of Hawaii,
640 North A`oh$\bar{o}$k$\bar{u}$ Place, Hilo, HI 96720, USA}
\author{Frantz Martinache}
\affiliation{Universit\'{e} C\^{o}te d'Azur, Observatoire de la C\^{o}te d'Azur, CNRS, Laboratoire Lagrange, France}
\author{Carol Grady}
\affiliation{Eureka Scientific, 2452 Delmer Street Suite 100, Oakland, CA, USA}
\affiliation{NASA-Goddard Space Flight Center, Greenbelt, MD, USA}
\author{Masa Hayashi}
\affiliation{National Astronomical Observatory of Japan,
Osawa 2-21-1, Mitaka, Tokyo 181-8588, Japan}
\author{Jungmi Kwon}
\affiliation{Institute of Space and Astronautical Science, Japan Aerospace Exploration Agency, 3-1-1 Yoshinodai, Chuo-ku, Sagamihara, Kanagawa 252-5210, Japan}
\author{Michael W. McElwain}
\affiliation{NASA-Goddard Space Flight Center, Greenbelt, MD, USA}
\author{Yi Yang}
\affiliation{Department of Astronomy, The Graduate University for Advanced Studies (SOKENDAI), National Astronomical Observatory of Japan, Japan}
\author{Motohide Tamura}
\affiliation{Astrobiology Center of NINS, 2-21-1, Osawa, Mitaka, Tokyo, 181-8588, Japan}


\begin{abstract}
We present new, near-infrared (1.1--2.4 $\micron$) high-contrast imaging of the bright debris disk surrounding HIP 79977 with the Subaru Coronagraphic Extreme Adaptive Optics system (SCExAO) coupled with the CHARIS integral field spectrograph.  SCExAO/CHARIS resolves the disk down to smaller angular separations of (0\farcs{}11; $r \sim 14$ au) and at a higher significance than previously achieved at the same wavelengths. The disk exhibits a marginally significant east-west brightness asymmetry in $H$ band that requires confirmation.
Geometrical modeling suggests a nearly edge-on disk viewed at a position angle of $\sim$ 114.6$\arcdeg$ east of north.
The disk is best-fit by scattered-light models assuming
strongly forward-scattering grains ($g$ $\sim$ 0.5--0.65) confined to a torus with a peak density at $r_{0}$ $\sim$ 53--75 au. We find that a shallow outer density power law of $\alpha_{out}=$-1-- -3
and flare index of $\beta = 1$ are preferred. Other disk parameters (e.g.~inner density power law and vertical scale height) are more poorly constrained.
The disk has a slightly blue intrinsic color and its profile is broadly consistent with predictions from birth ring models applied to other debris disks. 
While HIP 79977's disk appears to be more strongly forward-scattering than most resolved disks surrounding 5--30 Myr-old stars, this difference may be due to observational biases favoring forward-scattering models for inclined disks vs. lower inclination, ostensibly neutral-scattering disks like HR 4796A's.
Deeper, higher signal-to-noise SCExAO/CHARIS data 
can better constrain the disk's dust composition.
\end{abstract}

\section{Introduction} \label{sec:intro}
Debris disks around young stars are signposts of massive planets \citep[e.g.][]{Marois2008a,Lagrange2010} 
and critical reference points for understanding the structure, chemistry, and evolution of the Kuiper belt \citep{Wyatt2008}. 
Debris disks may be made visible by recently-formed icy Pluto-sized objects stirring and causing collisions between surrounding boulder-sized icy planetesimals.   The luminosity
distribution of debris disks over a range of ages then traces the evolution of debris produced by icy planet formation.
\citep{Currie2008, Kenyon2008}).   Similarly, massive jovian planets  
may create gaps in some of these debris disks and sculpt the distribution of their icy planetesimals \citep{Mustill2009}.   

Resolved imaging of debris disks in scattered light has revealed dust sculpted in morphologies ranging from diffuse structures or 
extended torii to sharp rings;
disks exhibited scattering properties ranging from neutral to strongly forward 
scattering \citep[e.g.][]{SmithTerrile1984,Schneider1999,Schneider2005,Schneider2009,Kalas2005,Kalas2006,Kalas2007hd15745,Soummer2014, Currie2015a, Currie2017a}.  Furthermore, multi-wavelength imaging and spectroscopy of debris disks in scattered light provide further insights into the nature of debris disk properties.   The differing grain properties of debris disks can result in a spread in intrinsic disk colors from blue ~\citep[e.g. AU Mic,][]{Fitzgerald2007}, where dust is reflecting light more efficiently at shorter wavelengths compared to what it receives from the star, to red~\citep[e.g. $\beta$ Pic,][]{Golimowski2006}. Detailed photometric color characterization provides insights into grain properties, and low-resolution spectroscopy (even as low as R $\sim$ 10) probes the presence of ices and organics \citep[e.g.][]{Debes2008,Rodigas2014,Currie2015a}.  

Extreme adaptive optics (ExAO) systems coupled with integral field spectrographs improve the ability to detect and characterize debris disks, especially at small angles.   For example, resolved imaging and spectroscopy of the HD 115600 debris disk with the \textit{Gemini Planet Imager}, the first object discovered with ExAO, revealed a sharp ring at $r$ $\lesssim$ 0\farcs{}5, modeling for which suggested neutral-scattering and possibly icy dust and a pericenter offset caused by a hidden jovian planet \citep{Currie2015a}.    \citet{Milli2017} resolved the well-known HR 4796A disk at far smaller angular separations than done previously.  They showed that a seemingly neutral-scattering dust ring has a strong forward-scattering peak at small angles, inconsistent with a single Henyey-Greenstein-like scattering function.  Resolved imaging and spectroscopy over a longer wavelength baseline enables better constraints on the properties of other debris disks \citep[e.g.][]{Rodigas2015, Milli2017}.

HIP 79977 is another young star whose debris disk can better understood using multi-wavelength imaging and spectroscopy with ExAO.  This is an F2/3V star (1.5 $M_{\sun}$) located $131.5 \pm 0.9$ pc away~\citep{Gaia2018} in the $\sim$ 10 Myr-old
Upper Scorpius association \citep{Pecaut2012}. Its infrared excess was detected by the $IRAS$ satellite, and the Spitzer Multiband Imaging Photometer associated it with a debris disk~\citep{Chen2011}. \citet{Thalmann2013} used Subaru's facility (conventional) AO188 adaptive optics system and the HiCIAO instrument at $H$ band and produced the first resolved images of its debris disk. They revealed that it was
viewed nearly edge-on 
($i=84^{+2}_{-3}\arcdeg$) and had a position angle of $\rm{PA} = 114.0 \arcdeg \pm 0.3 \arcdeg$.
The noted tangential linear polarization varying from $\sim$ 10$\%$ at 0\farcs{}5 to $\sim$ 45$\%$ at
1\farcs{}5. \citet{Engler2017} performed the first ExAO characterization of HIP 79977, observing it at visible 
wavelengths ($\lambda_c = 735$ nm, $\Delta \lambda = 290$ nm) using the SPHERE-ZIMPOL polarimeter. They
measured a polarized flux contrast ratio for the disk of 
$(F_{\rm pol})_{\rm disk}/F_{\star} = (5.5 \pm 0.9) \times 10^{-4}$ in that band and an increase in the thickness
of the disk at larger radii, which they explained by the blow-out of small grains by stellar winds. They found a best-fitting
inclination of $i=84.6 \arcdeg \pm 1.7 \arcdeg$ and a position angle of $\rm{PA} = 114.5 \arcdeg \pm 0.6 \arcdeg$.

These previous studies showed tension in some derived debris disk properties (e.g. the disk radius) and allowed a wide range of parameter space for others (e.g. the disk scattering properties).  No substellar companions were decisively detected in either publication.  However, \citet{Thalmann2013} did find a marginally significant point-like residual emission in their reduced image after subtracting a model of the debris disk's emission.

In this paper, we present the first near-IR resolved ExAO images of the HIP 79977 debris disk, using the Subaru Coronagraphic Extreme Adaptive Optics systems coupled with the CHARIS integral field spectrosgraph.   SCExAO/CHARIS data probe  inner working angles (0\farcs{}15--0\farcs{}2) comparable to those from SPHERE polarimetry reported in~\citet{Engler2017} and significantly smaller than that presented in~\citet{Thalmann2013}. Additionally, we present the first near-IR color analysis of the disk. 

The paper is 
organized as follows: in Section~\ref{sec:obs}, we describe the observations and the pipeline through
which the data was reduced and then PSF-subtracted; in Section~\ref{sec:geometry}, we describe the
basic morphology of the disk; then, in Section~\ref{sec:methodology} we discuss the process through which
we generated synthetic disks and propagated them through the same pipeline as the actual data in order 
to understand how the PSF subtraction attenuated the disk features; we provide the results of this
forward modeling in Section~\ref{sec:results}; finally, we describe the $J$-, $H$-, and $K_p$-band colors of the 
disk.

\section{SCExAO/CHARIS Data}\label{sec:obs}
\subsection{Observations and Data Reduction}
We targeted HIP 79977 on UT 14 August 2017 (Program ID S17B-093, PI T. Currie)
with Subaru Telescope's SCExAO~\citep{Jovanovic2015} instrument coupled
to the CHARIS integral field spectrograph, which operated in low-resolution ($R\sim 20$), 
broadband (1.13--2.39 $\micron$) mode \citep{Peters2012,Groff2013}.
SCExAO/CHARIS data were obtained using the Lyot coronagraph with the 217 mas diameter occulting spot. 
Satellite spots, attenuated copies of the stellar PSF, were generated by placing
a checkerboard pattern on the deformable mirror with a 50 nm amplitude and alternating its
phase between $0\arcdeg$ and $180\arcdeg$~\citep{Jovanovic2015b}. These spots were used for image
registration and spectrophotometric
calibration; their intensity relative to the star\footnote{The spot intensity calibration
changed following the observations described in this paper, so this equation may
not match what is provided elsewhere.} was given by
\begin{equation}
I_{spots}/I_{\star} = 4 \times 10^{-3} (\lambda/1.55 \; \micron)^{-2} .
\label{eqn:satspots}
\end{equation}
Exposures consisted of 86 co-added 60 $s$
frames (82 science frames, 4 sky frames) obtained in pupil tracking/angular differential imaging \citep[ADI,][]{Marois2006} mode over 92 minutes and covering a total parallactic angle rotation of $26.7\arcdeg$. Conditions were excellent; seeing was 0\farcs{}35-0\farcs{}40 at 0.5 $\micron$ and the wind speed was 3 m s$^{-1}$.
Although we did not obtain a real-time estimate of the Strehl ratio, the raw contrasts at $r$ $\sim$ 0\farcs{}2--0\farcs{}75 later estimated from spectrophotometrically calibrated data were characteristic of those 
obtained with $H$-band Strehls of 70--80\% \citep{Currie2018a}.  

 \begin{figure*}[!ht]
 \begin{centering}
 \includegraphics[width=14cm]{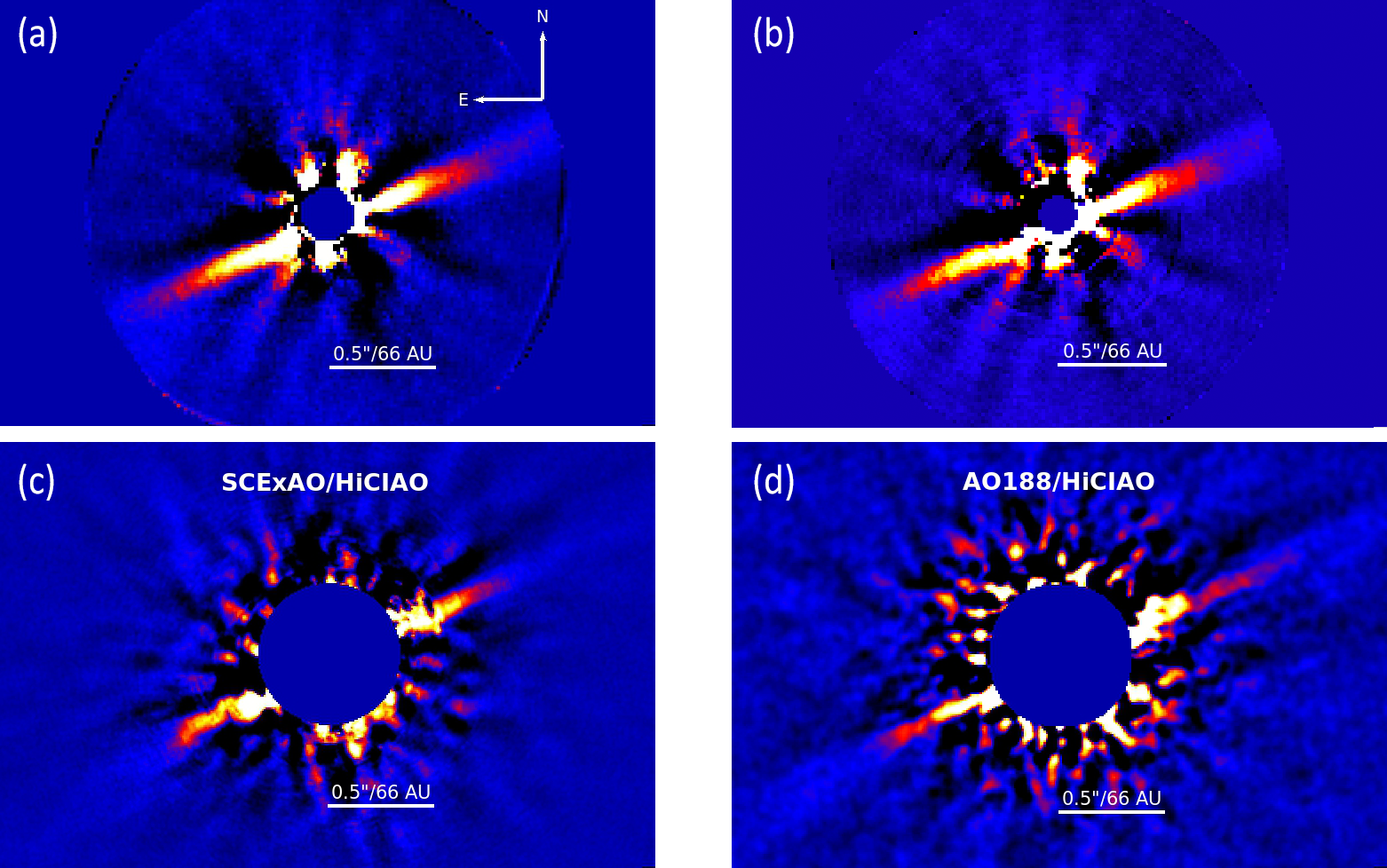}
 \caption
 {Illustrated here are the three different NIR datasets for HIP 79977.  The upper panels (a and b) are our 
 paper's main focus and
 show wavelength-collapsed images produced by two different KLIP-ADI reductions of the SCExAO/CHARIS data.
 Figure~\ref{fig:diskcomparison}c shows July 2016 $H$ band data from SCExAO + HiCIAO reduced using A-LOCI with local masking~\citep{Currie2012} 
 and has stronger residuals exterior to 0\farcs{}3--0\farcs{}4. 
 Finally, Figure~\ref{fig:diskcomparison}d shows the data published by~\citet{Thalmann2013},
 which were produced using the (non-extreme) AO188 and HiCIAO at $H$ band.  The data were processed using the ACORNS-ADI reduction package~\citep{Brandt2013}.
 The four images have the same intensity scaling. The circular region in the bottom
 two plots denotes the field of view of the CHARIS data.
 }
 \label{fig:diskcomparison}
 \end{centering}
 \end{figure*}

To convert raw CHARIS files into data cubes, we employed the CHARIS Data Reduction Pipeline \citep[CHARIS DRP,][]{Brandt2017}.  After generating a wavelength solution from monochromatic ($\lambda_{0}$ = 1.550 $\micron$) lenslet flats, the pipeline extracted data cubes using the least squares method described by \citet{Brandt2017}, yielding a nominal spaxel scale of 0\farcs{}0164 and $\sim$ 1\farcs{}05 radius field of view.  
Subsequent processing steps -- e.g. image registration and spectrophotoemtric calibration -- followed those from \citet{Currie2018a}.

For PSF subtraction, we utilized the Karhunen-Lo\`eve Image Projection (KLIP)-based algorithm of \citet{Soummer2012} in angular differential imaging-only mode as employed in 
\citet{Currie2014hd100546,Currie2017b}, where PSF subtraction is performed in annular regions with a 
rotation gap to limit signal loss from self-subtraction of astrophysical sources.   Key algorithm 
parameters -- the width of annulus over which PSF subtraction is performed ($\Delta$r), the rotation gap ($\delta$), the number of principal components ($N_{\rm pc}$) -- were varied to explore which combination maximized the total SNR of the disk in sequence-combined, wavelength-collapsed images.  While the detection of the HIP 79977 debris disk was robust across the entire range of parameter space, the signal to noise of the spine of the disk was maximized with a setting with $\Delta$r = 2 pixels, $N_{\rm pc} =2$, and $\delta = 1.0$ full width half maxima (FWHM) and then merging
the wavelength channels using a robust mean with outlier rejection instead of a median combination.  As described later, for computational efficiency and simplicity, we performed a second reduction with a larger annular width of $\Delta$r = 6 pixels ($\sim 2.5 \lambda$/D at 1.55 $\micron$).  Reductions retaining a slightly different number of principal components or value for the rotation gap yielded comparable results.  
 
\subsection{Detection of the HIP 79977 Debris Disk}
Figures~\ref{fig:diskcomparison}a and~\ref{fig:diskcomparison}b show the results of these two reductions of the
CHARIS data. Figures~\ref{fig:diskcomparison}c and~\ref{fig:diskcomparison}d contextualize the performance gain of SCExAO/CHARIS compared to earlier
observations. The disk is plainly visible down to an inner working angle of 
0\farcs{}11 in~\ref{fig:diskcomparison}a and~\ref{fig:diskcomparison}b.
Figure~\ref{fig:diskcomparison}c shows data collected on UT 17 July 2016 (Program UH-12B, PI K. Hodapp) using SCExAO (suboptimally tuned
and providing lower Strehl than that of the recent data) and the HiCIAO instrument at $H$ band.
Although the July 2016 SCExAO/HiCIAO image has a larger
field of view than the SCExAO/CHARIS image, it exhibits far stronger residuals interior to about 0\farcs{}3--0\farcs{}5.  
Figure~\ref{fig:diskcomparison}d shows the AO188~\citep[Subaru's facility AO system,][]{minowa10} + HiCIAO data previously published
by~\citet{Thalmann2013}, and this has even stronger residuals, particularly at smaller angular separations,
due to its much poorer AO correction.

\begin{figure*}[!ht]
 \begin{centering}
 \includegraphics[width=1.0\textwidth]{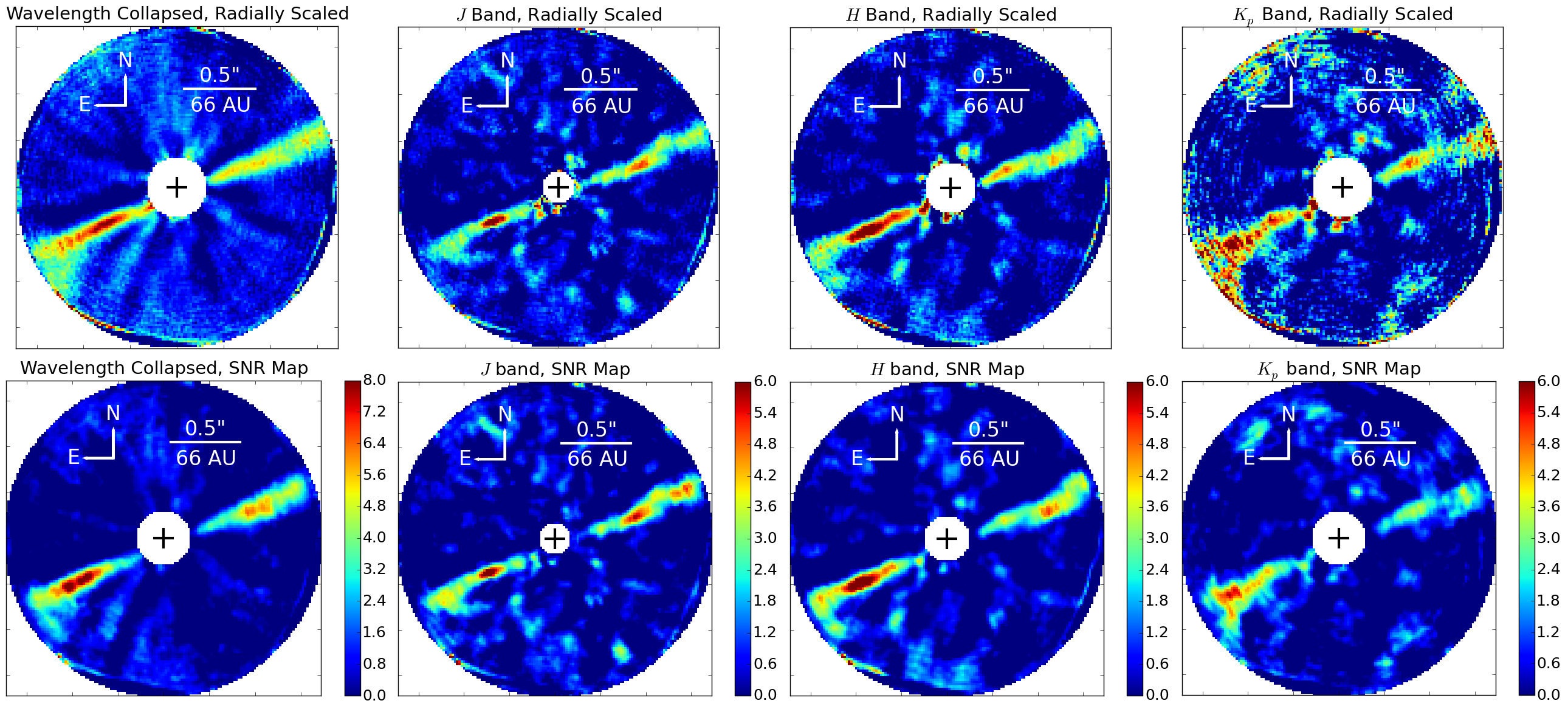}
 \caption
 {Shown here are flux images following KLIP PSF-subtraction (top) and the corresponding 
 signal-to-noise per resolution element
 maps (bottom). The CHARIS low-resolution mode produces data cubes with 22 spectral layers. We coadded all the layers (left) and the bands corresponding to (proceeding rightward) $J$, $H$, and $K_p$ bands. The flux images have
 arbitrary units and have been multiplied by an $r^2$ map in order to reveal structure away from the star.   
 The images presented here are rotated relative to those in Figure~\ref{fig:diskcomparison}.}
 \label{fig:combineddiskimages}
 \end{centering}
 \end{figure*}

Figure~\ref{fig:combineddiskimages} shows the sequence-combined, wavelength-collapsed disk image scaled by the stellocentric distance squared, and analogous images obtained from combining channels covering the $J$ (channels 1--5; 1.16--1.33 $\mu$m), $H$ (channels 8--14; 1.47--1.80 $\mu$m), and $K_{p}$ (channels 16--21; 1.93--2.29$\mu$m) passbands. This image used the first set of KLIP parameters described above.
The disk is plainly visible in each image.  We computed the signal-to-noise per resolution element using the standard practice of replacing each pixel with the sum within a FWHM-sized aperture, computing the radial profile of the robust standard deviation of this summed image in the wavelength-collapsed image, dividing the two images, and correcting for small sample statistics \citep{Currie2011a}.   The disk is detected at a SNR/resolution element (SNRE) $>$ 3 exterior to 0\farcs{}25 and peaks at SNRE $\sim$ 9.1, 8, 9.1, and 5.8 in the broadband, $J$, $H$, and $K_{p}$ images, respectively\footnote{We achieved comparable results using a different algorithm, A-LOCI, using local masking as implemented in \citet{Currie2012,Currie2017a}.}.   These estimates are conservative as we do not mask the disk signal when computing the noise profile.  For our second reduction the SNRE values along the disk spine are slightly smaller at small angles but otherwise comparable, peaking at 9.6, 9, 8.4, and 5.6 in the broadband, $J$, $H$, and $K_{p}$ images, respectively.
 
For both reductions, the final images and SNR maps may reveal some evidence for a wavelength dependent brightness asymmetry between the eastern and western sides.  In the wavelength-collapsed image, the eastern
side of the disk appears about 50\% brighter and is detected at a higher significance ($\sim 8-9$ $\sigma$ vs. 
$5.5-6.5$ $\sigma$ along the disk spine beyond 0\farcs{}5).   From comparing images obtained over different passbands,  $H$ and $K_{p}$ band seem to be responsible for most of this brightness asymmetry.   

\section{Geometry of the HIP 79977 Debris Disk}\label{sec:geometry}
Our images clearly trace the major axis of the HIP 79977 debris disk.   To estimate the disk's position angle, we follow previous analysis performed for HD 36546 \citep{Currie2017a} and for $\beta$ Pic \citep{Lagrange2012}, determining the trace of the disk spine from the peak brightness as a function of separation (``maximum spine'' fitting) and from fitting a Lorentzian profile.   Our procedure used the \textit{mpfitellipse} package to estimate the disk spine from disk regions between 0\farcs{}15 and 0\farcs{}75, where the pixels are weighted by their SNRE, and explored a range of thresholds in SNRE (0--3) to define the spine.   

Precise astrometric calibration for CHARIS is ongoing and preliminary results will be described in full in a separate early-science paper focused on $\kappa$ Andromedae b (Currie et al. 2018, in prep.).   Briefly, we obtained near-infrared data for HD 1160 from SCExAO/CHARIS in September 2017 and Keck/NIRC2 in December 2017.  At a projected separation of $r$ $\sim$ 80 au, the low-mass companion HD 1160 B should not experience significant orbital motion \citep{Nielsen2012, Garcia2017}; Keck/NIRC2 is precisely calibrated, with a north position angle uncertainty of 0.02$^{o}$ and post-distortion corrected astrometric uncertainty of 0.5 mas \citep{Service2016}.   Thus, we pinned the SCExAO/CHARIS astrometry for HD 1160 B to that for Keck/NIRC2 to calibrate CHARIS's pixel scale and north position angle offset.   These steps yielded a north PA offset of $\sim -2.2\arcdeg$ east of north and a revised pixel scale of $\sim$ 0\farcs{}0162.  While the differences between the default and revised pixel scale lead to astrometric offsets are inconsequential for this paper (10 mas near the edge of CHARIS's field of view), the north position angle (PA) offset for CHARIS is necessary for an accurate estimate of the position angle for the disk's major axis.

After considering CHARIS's north PA offset, 
Lorentzian profile fitting yields a position angle of $114.59\arcdeg \pm 0.40\arcdeg$.   ``Maximum spine" fitting yields nearly identical results but with larger error bars:
$114.74\arcdeg \pm 1.88\arcdeg$.   These values are consistent with previous estimates from \citet{Engler2017} and \citet{Thalmann2013}.   For the rest of the paper, we adopt a position angle of $114.6\arcdeg$. 

\section{Modeling of the HIP 79977 Debris Disk}\label{sec:forwardmodeling}
\subsection{Methodology}\label{sec:methodology}
\subsubsection{Forward-Modeling of the Annealed Disk Due to PSF Subtraction}
To assess the morphology of the HIP 79977 debris disk, we forward-modeled synthetic disk images spanning a range of properties through empty data cubes, using the same eigenvalues and eigenvectors used in the reduction of our on-sky data \citep[e.g.][]{Soummer2012,Pueyo2016}.  Our specific implementation, following the formalism in \citet{Pueyo2016}, is described and justified in detail below.

The residual signal of a planet or disk in a target image with spatial dimensions x and an intrinsic signal \textbf{A(x)} after KLIP processing is nominally equal to the astrophysical signal in the target image minus its projection on the KLIP basis set constructed from references images from up to k = 1 $\cdots$ $K_{\rm klip}$ principal components, \textit{$Z_{k}$}:
\begin{equation}
P_{\rm residual,n} = A(x_{n}) - \Big(\sum_{k=1}^{K_{\rm klip}}<A(x_{n}),Z_{k}^{KL}> Z_{k}^{KL}(n)\Big)
\end{equation}

Here, \textbf{Z$_{k}$$^{KL}$} is the Karhunen-Loe\'ve transform of the reference image library \textbf{R} 
with eigenvalues \textbf{$\Lambda_{k}$} and eigenvectors
\textbf{$\nu_{k}$}:

\begin{equation}
Z_{k}^{KL}(x) = \frac{1}{\sqrt{\Lambda_{k}}}\sum_{m=1}^{K_{\rm klip}}\nu_{k}R_{m}(x).
\end{equation} 

When the astrophysical signal in a given image is not contained in reference images used for subtraction or is negligible, then annealing is  due to \textit{oversubtraction} -- confusion of the astrophysical signal with speckles -- and is fully described by a straightforward application of Equation 2.
As described in \citet{Pueyo2016}, however, the presence of an astrophysical signal in the reference image library itself causes \textit{self-subtraction} of the source in the target image and perturbs \textbf{Z$_{k}$$^{KL}$} by an amount \textbf{$\Delta Z_{k}$$^{KL}$}.   
Self-subtraction can further be subdivided into two contributions.  \textit{\underline{Direct} self-subtraction} scales linearly with the astrophysical signal ($\epsilon$) and inversely with the square-root of the unperturbed eigenvalues: \textbf{$\Delta Z_{k}$$^{KL}$} $\propto$ $\epsilon$/$\sqrt{\Lambda_{k}}$. \textit{\underline{Indirect} self-subtraction} is inversely proportional to the eigenvalues: \textbf{$\Delta Z_{k}$$^{KL}$} $\propto$ $\epsilon$/${\Lambda_{k}}$.

\citet{Pueyo2016} qualitatively discuss the typical cases where oversubtraction and the two different types of self-subtraction (direct and indirect) dominate for point sources.
For small $K_{\rm klip}$ values and an astrophysical signal that is small compared to the speckles over the region where principal component analysis is performed, oversubtraction usually is the primary source of annealing.
For intermediate $K_{\rm klip}$ values, \textit{\underline{direct} self-subtraction} usually dominates.  For large $K_{\rm klip}$, closer to a full-rank covariance matrix, \textit{\underline{indirect} self-subtraction} becomes the most important term.
 However, the relative contribution of each of these terms for a given $K_{\rm klip}$ value depends on the nature of the astrophysical source  to be detected (e.g. planet, sharply defined disk, diffuse disk) and other algorithm settings.   For example, using a larger rotation gap can remove more astrophysical signal from the reference library, reducing the influence of self-subtraction at a given $K_{\rm klip}$.

Previous measurements of the HIP 79977 disk further help identify the important biases/sources of annealing for our HIP 79977 data set.  In our reductions, the number of removed KL modes (2) is small compared to the size of the reference library 
(82 $N_{\rm images/channel}$).  In most channels, the disk is $\approx$ 5\% of the brightness of the local speckles.  Furthermore, we perform PSF subtraction in annular regions.  Over the angular separations modeled (0\farcs{}16--0\farcs{}75), results from \citet[][see their Fig. 6b]{Engler2017} imply that the nearly edge-on disk is present in no more than 20\% of the pixels at each angular separation.    Our rotation gap criterion (1 PSF footprint) further reduces self-subtraction.   As a result, the perturbed KL modes $\Delta$KL are far smaller than the unperturbed ones dominated by signal from the speckles: the indirect self-subtraction term is negligible.
Thus, in performing forward-modeling we consider oversubtraction and direct self-subtraction only.   

\subsubsection{Scattered Light Disk Models}
\begin{figure}[!ht]
 \begin{centering}
 \includegraphics[width=0.4\textwidth]{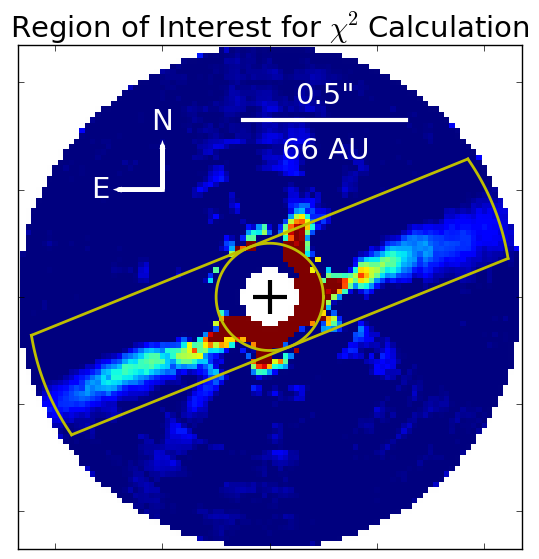}
 \caption
 {The region bounded by the yellow lines was used for scaling the PSF-subtracted
 synthetic model disks and then computing their $\chi^2$ residuals relative to the
 on-sky data. The outer boundary is defined by the intersection of a rectangular box that
 is 100 pixels by 20 pixels where the major axis is rotated $22\arcdeg$ north of west and 
 a circle of radius $r=45$ pixels. The inner region is a circle of radius $r=10$ pixels.
 The disk in this figure is plotted from the same data as that used in 
 Figure~\ref{fig:combineddiskimages}, but it 
 has not been multiplied by an $r^2$ map.}
 \label{fig:roi}
 \end{centering}
 \end{figure}

Synthetic scattered light disk models were drawn from the GRaTeR code developed in \citet{Augereau1999}, convolved with the SCExAO/CHARIS instrumental PSF, and inserted into empty data cubes with the same position angles as the real data.  We then forward-modeled the annealing of each model disk in each wavelength channel due to KLIP PSF subtraction as described above and compared the wavelength-collapsed image of the residual disk model to the real data.  The fidelity of each model disk to the data is determined in the subtraction residuals binned (by the instrument PSF size of $\sim$ 0\farcs{}04, which corresponds to the area of 7 pre-binned pixels) over a region of interest defining the trace of the disk and any self-subtraction footprints (see Figure~\ref{fig:roi}).
This evaluation region encloses 237 binned pixels (N$_{\rm data}$). 
 
The set of acceptably-fitting solutions have chi-squared values of 
$\chi^{2} \le \chi^{2}_{\rm min} + \sqrt{2 N_{\rm data}}$ \citep[see][]{Thalmann2013}.  
At the 95\% confidence limit, this criterion equals $\chi_{\nu}^{2} \lesssim 1.092$. 

Because we performed KLIP PSF subtraction in annuli (not the entire field at once) and in each of the 22 wavelength channels separately (not single-channel camera data), exploring 10$^{6}$ models covering a large parameter space as in \citet{Engler2017} would be extremely computationally expensive.  Rather, we leverage on inspection of the SCExAO/CHARIS wavelength-collapsed final image, our disk geometry modeling, and previous results from \citet{Engler2017} to focus on a smaller parameter space range.   

Inspection of the final CHARIS image shows that the disk is detected only on the near side, out to an angular 
separation of 0\farcs{}5--0\farcs{}6 before gradually fading in brightness at wider separations.   
Our fitting to the geometry of the disk reaffirms the position angle of 114.6$^{o}$ we adopted in Section~\ref{sec:geometry}.  Thus, our parameter space 
generally explores disks with moderate to strong forward-scattering, a sharp inner cutoff to the belt, and a shallower decay in dust density beyond the fiducial radius.

We varied six parameters in our search for the disk that best reproduced the on-sky data.
First, the Henyey-Greenstein parameter~\citep{Henyey1941} probes the visible extent of the dust's phase scattering function.
While it lacks a pure physical motivation and is known to fail at very small scattering angles for at least some debris disks \citep{Milli2017}\footnote{These angles correspond to the semiminor axis of the HIP 79977 debris disk,  which is too close to the star to be accessible with our data.}, it is widely adopted in debris disk modeling literature and thus helps cast our results within the context established by other debris disks.   The H-G parameter ranges from $-1$ to $1$;
$g=0$ corresponds to neutral scattering, $g=-1$ indicates perfect backward scattering, and $g=1$ 
indicates dust that scatters light solely forward.

Second, we varied values of the fiducial radius $r_0$ of the disk, inside of which $\alpha_{in}$ ($\alpha_{in} > 0$)
describes the power law for the increase in dust particle number density and outside
of which $\alpha_{out}$ ($\alpha_{out} < 0$) describes the power law for its decrease.
These three variables, which were the second through fourth fitted parameters,
combine to give the radial distribution profile $R(r)$:
\begin{equation}
R(r) = \left[ \Big( \frac{r}{r_0} \Big) ^{-2 \alpha_{in}} + 
  \Big( \frac{r}{r_0} \Big)^{-2 \alpha_{out}} \right]^{-1/2}
\end{equation}
where $r$ is the distance from the center of the disk. The vertical profile $Z(h)$ is given by
\begin{equation}
Z(h) = \textrm{exp} \left( \frac{-| h |}{H(r)} \right) ,
\end{equation}
where
$h$ is the distance above the disk midplane. $H(r)$ is the scale height at radius $r$ and is given by
\begin{equation}
H(r) = \xi \left( \frac{r}{r_0} \right)^\beta ,
\end{equation}
where $\xi$ is the scale height at $r_0$ and $\beta$ is the disk's flare index. $\xi$ and $\beta$ were
the fifth and sixth parameters in our grid search.

We tested models with $g=0.3-0.8$, corresponding to moderate to strong forward scattering. 
Based on visual estimates of the disk images,
we produced model disks with fiducial radii of $r_0 = 43-91$ au. The parameters $\alpha_{in}$ and $\alpha_{out}$ determine
the power laws for the inner and outer radial emission profiles, respectively, and we selected values that produced
disks with relatively sharp inner cutoffs and slow radial decays. We sampled disks with a scale height at
the fiducial radius in the range of $\xi=0.5-3.2$ au; values outside this range would not be consistent with the
self-subtracted images. 
We adopt our value for the disk position angle determined in Section~\ref{sec:geometry}.
and 
used our available computing resources to probe a greater variety of the other parameters.
Values outside these parameter ranges produced synthetic disks whose morphology differed greatly 
from the on-sky results.
The left two columns of Table~\ref{table:syndisks} list each parameter and the associated range in 
parameter space explored. A total of 20,480 disks were considered.

Our nominal search considered only circular disks with no star offset, which was the same approach taken
by~\citet{Engler2017}. As stated previously, because the position angle
and inclination were tightly constrained by~\citet{Engler2017} and~\citet{Thalmann2013} and our spine fitting reaffirmed
their values, we adopted a position angle of $114.6\arcdeg$ and an inclination of $i=84.6\arcdeg$. 
\citet{LiemanSifry2016} analyzed ALMA data and also measured a position angle and inclination consistent with this.

\begin{deluxetable*}{lccc}[!ht]
\tablecaption
{The grid of synthetic model disks used in our forward modeling. \label{table:syndisks}}
\tablecolumns{4}
\tablenum{1}
\tablewidth{0pt}
\tablehead{
\colhead{Parameter} & \colhead{Values tested} & \colhead{Value for} & \colhead{Acceptably-fitting} 
\\
\colhead{} & \colhead{} & \colhead{best model} & \colhead{values} 
}
\startdata
 Radius of belt $r_0$ (au) & [43, 53, 64, 69, 75, 80, 86, 91]  & 53 & [53, 64, 69, 75] \\
 Inner radial index $\alpha_{in}$ & [3, 4, 5, 6] & 6 & [3, 4, 5, 6]   \\
 Outer radial index $\alpha_{out}$ & [-1, -1.5, -2, -2.5, -3, -3.5, -4.5, -5.5] & -1.5 & [-1, -1.5, -2, -2.5, -3]  \\
 Vertical scale height $\xi$ (au) & [0.5, 1.1, 1.6, 2.1, 3.2] & 3.2 & [0.5, 1.1, 1.6, 2.1, 3.2] \\
 Flare index $\beta$ & [1, 2] & 1 & [1, 2]   \\
 H-G parameter $g_{sca}$ & [0.3, 0.4, 0.5, 0.55, 0.6, 0.65, 0.7, 0.8] & 0.6 & [0.5, 0.55, 0.6, 0.65] \\
\enddata
\tablecomments{We adopted values for inclination $i=84.6\arcdeg$, eccentricity $e=0$, and position angle $
\theta=114.6\arcdeg$ in accordance with those measured by~\citet{Engler2017}. $\xi$ and $r_0$ are not
round numbers because they were initially chosen based
on the distance to HIP 79977 provided by~\citet{vanLeeuwen2007}, which was refined by~\citet{Gaia2018}, 
causing the scale to change by $\sim 7 \%$. If one value of a parameter fell below the
acceptably-fitting $\chi_{\nu}^{2}$ threshold for at least one model, it was included here.
Figure~\ref{fig:histograms} shows which parameters values most frequently produced acceptably-fitting
models.}
\end{deluxetable*}

\subsection{Results}\label{sec:results}
\begin{figure}[!ht]
 \begin{centering}
 \includegraphics[width=0.47\textwidth]{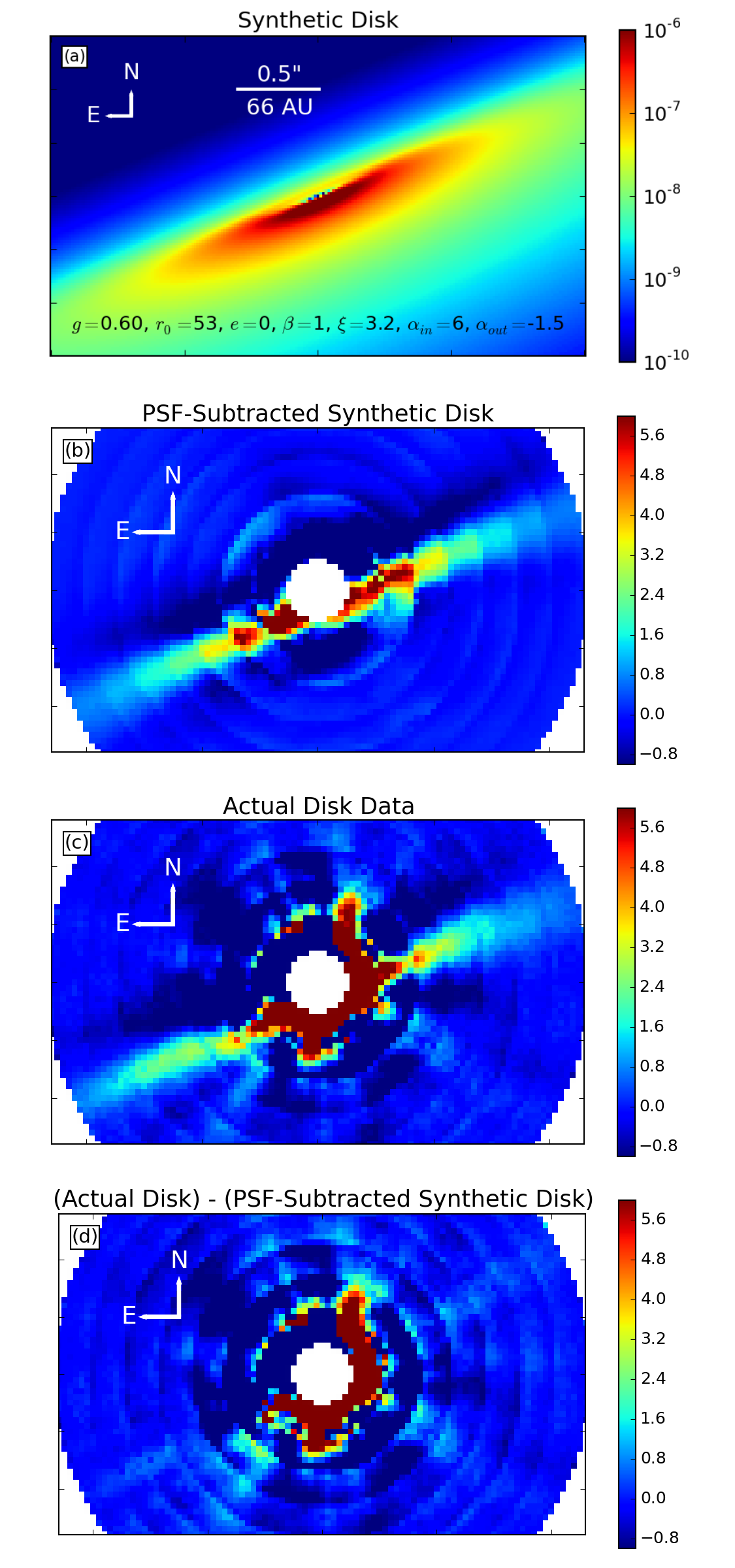}
 \caption
 {From top to bottom are (a) the best-fitting synthetic disk; (b) that disk after
 it was convolved with the SCExAO PSF and then propagated through the KLIP PSF-subtraction
 using the same eigenvalues and eigenvectors as the on-sky data; (c) the wavelength-collapsed
 disk image (same as Figure~\ref{fig:diskcomparison}b) used in the $\chi^2$ comparison with 
 the synthetic model; and (d) the difference between panels (c) and (b). Minimal structure 
 remains in panel (d), indicating that the synthetic disk closely matches the actual data.
 The units are arbitrary. The distance scale is the same in all four panels.}
 \label{fig:bestsyndisk}
 \end{centering}
\end{figure}

Of the 20,480 synthetic disks, 132 produced residuals of $\chi^{2}_{\nu} \lesssim 1.092$ and 
therefore were acceptably fitting. 
The best model, which we defined as the model yielding the smallest 
$\chi^2_{\nu}$, produced $\chi^{2}_{\nu}$ $\sim$ 1, suggesting that the best-fit models meaningfully reproduce the data.
The three panels of Figure~\ref{fig:bestsyndisk} show the best-fitting synthetic disk before 
and after PSF subtraction and the resulting residuals after it was subtracted from the on-sky 
data. This disk model
had $g=0.6$, indicating moderately strong forward scattering, a fiducial radius of $r=53$ au, a flare
index of $\beta=1$, a disk scale height at 
the fiducial radius of $\xi=3.2$ au, and dust emission with an inner power law of $\alpha_{in}=6$ 
and outer power law of $\alpha_{out}=-1.5$.

\begin{figure}[!ht]
\hspace*{-7mm}
 \begin{centering}
 \includegraphics[width=0.57\textwidth]{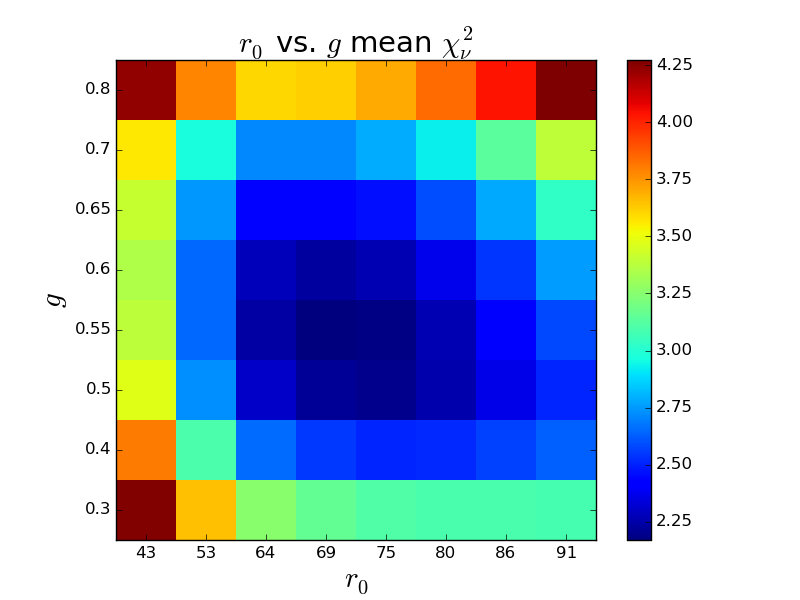}
 \caption
 {Shown here is the mean $\chi^2_{\nu}$ for each value of $r_0$ and $g$. All values
 for the other parameters were included in the mean when calculating the value of each pixel. 
 We produced
 these maps for every variable against every other variable; this map is illustrative of the
 results. We used these maps to verify that our tested values adequately spanned the
 parameter space.   The region of parameter space minimizing $\chi^{2}$ is clear and well behaved.}
 \label{fig:contourmap}
 \end{centering}
\end{figure}

\begin{figure}[!ht]
 \begin{centering} 
 \includegraphics[width=0.5\textwidth]{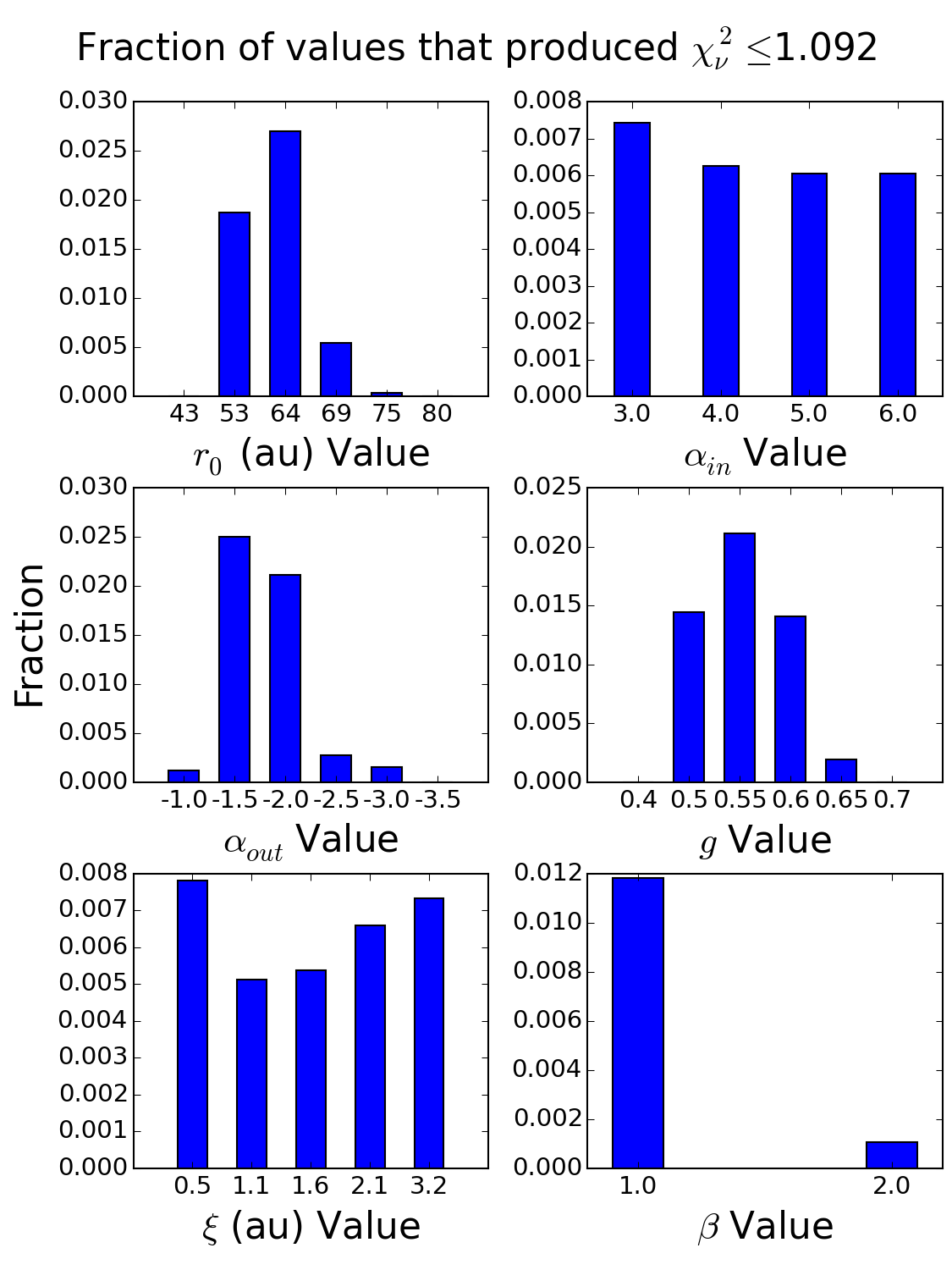}
 \caption
 {Each histogram bin contains the ratio of all models with that parameter value that
 produced an acceptably fitting $\chi_{\nu}^2$ compared to the number of models with that parameter
 value. The average of the bin heights in each plot is $132/20480 \approx 0.0064$. Some values with zero well-fitting disks have not been plotted in order to improve readability.}
 \label{fig:histograms}
 \end{centering}
\end{figure}

The range of parameters covered by 
the acceptably-fitting models is summarized in the fourth column of Table~\ref{table:syndisks}.  We 
produced contour maps of the average fit quality for every value of every parameter against every value of
every other parameter. An example map, showing the average $\chi^{2}_{\nu}$ for each value of $r_0$ and
$g$ averaged across the other parameters, is shown in Figure~\ref{fig:contourmap}. These maps helped
us ensure that we were sampling a reasonable range of values for each parameter.
Additionally, histograms of the parameter values that produced these acceptably-fitting models are shown
in Figure~\ref{fig:histograms}.  


Our modeling yielded improved constraints on the disk's radius and its scattering properties. As
shown in Figure~\ref{fig:contourmap}, there is a clear minimum in $\chi^2_{\nu}$ around
$g \approx 0.55$ and $r_0 \approx 64$ au. As shown in Figure~\ref{fig:histograms}, the family
of acceptably-fitting solutions has a small spread around these values. Our contour plots showed a 
strong preference for $\beta = 1$, indicating that the disk has low flaring.

On the other hand, the acceptably-fitting models covered the full range of considered
values of $\alpha_{in}$, indicating that $\alpha_{in}$ is not further constrained by our model fitting
beyond what was done in \citep{Engler2017}. This is likely because there was inadequate disk available
between the inner working angle and the fiducial radius for the $\alpha_{in}$ fitting to occur.

Our assumption that the disk is circular and has zero stellar offset is affirmed by the fact that
$\chi^{2}_{\nu} \sim 1$, which indicates that (within errors) the model accurately reproduces the data.
We did try forward modeling a small number of synthetic disks with low eccentricity or small stellar offset
but other parameters identical to those of the best-fitting disk, and the $\chi^{2}_{\nu}$
residuals were the same or slightly worse than those from the best-fitting circular disks.

We find numerically a good match between the wavelength-collapsed image and forward-modeled non-eccentric 
disk models, which show no brightness asymmetry. However, as evidenced by 
Figure~\ref{fig:combineddiskimages}, the HIP 79977 disk appears to exhibit asymmetrical
brightness. The east side of the disk is clearly brighter than the west side in $H$ band, 
and less clearly so in others. 
This brightness asymmetry may also be present
in SCExAO/HiCIAO $H$ band data from 2016 (Figure~\ref{fig:diskcomparison}c). 
This suggests that it may not be an artifact of the
data or processing. Plausible causes of the disk asymmetry are discussed in 
Section~\ref{sec:discussion}.

\section{HIP 79977 Disk Surface Brightness Profile and Colors}\label{sec:colors}
Next, we computed the surface brightness profile of the HIP 79977 disk in the $J$, $H$, and $K_p$ bands. We
began by using the satellite speckles (the PSF core was hidden by a coronagraph, but 
the flux of the satellite speckles was given by Equation~\ref{eqn:satspots}) and knowledge of the star's 
spectral type to spectrophotometrically calibrate the data cube. Second, we rotated the 
image so that the disk's spine was approximately horizontal and then fitted modified Gaussian
functions along the disk in order to find the spine's location with greater precision. We
then fit a fourth-order polynomial to these positions in order to smooth them and used
this fit as the trace of the disk in the subsequent steps.
Next, we merged the appropriate spectral channels to produce images equivalent to $J$, $H$, and $K_p$ bands 
and calculated a nominal surface brightness in each band along the disk's spine at radial intervals
of one PSF footprint. Uncertainties were calculated using the technique described in Section~\ref{sec:obs}. We 
divided the post-PSF-subtraction best-fitting synthetic model 
disk by the pre-PSF-subtraction version in order to produce a map of the attenuation that occurred
during the PSF subtraction. The PSF subtraction attenuated the disk spine by typically $25-40\%$,
and the attenuation increased with vertical displacement from the disk. Finally, we scaled the 
nominal surface brightnesses by to the attenuation map. 

\begin{figure}[!ht]
 \begin{centering}
 \includegraphics[width=0.45\textwidth]{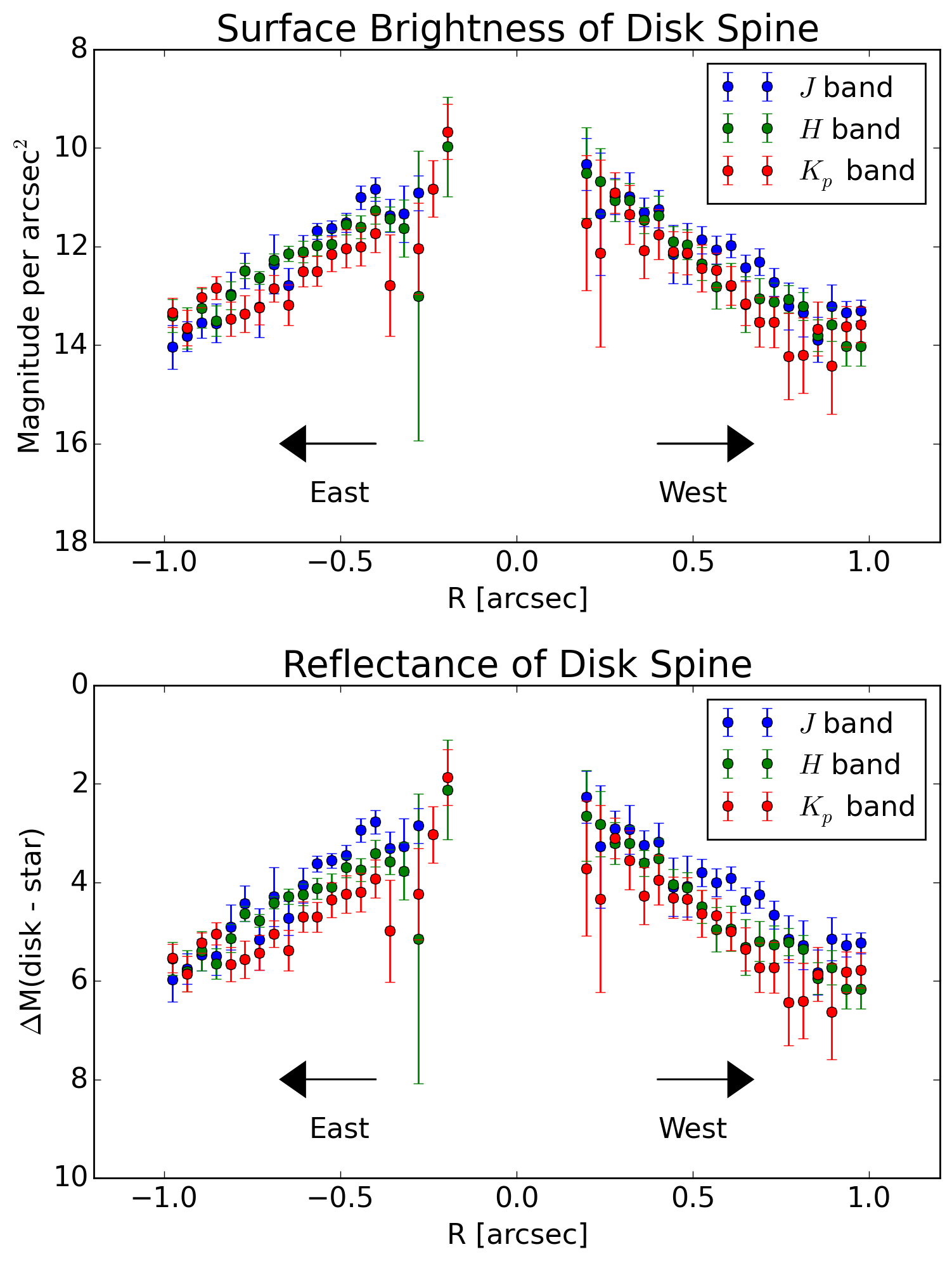}
 \caption
 {The $J$, $H$, and $K_p$ band surface brightnesses along the disk spine are shown in
 the top plot.
 In the lower plot, we have subtracted the flux of the star ($J=8.062, H=7.854, K=7.800$)
 in order to see the disk's colors after removal of the stellar color.
 The disk is slightly blue at most radial separations.
 The three bands plotted individually are shown in Figure~\ref{fig:sb2}.}
 \label{fig:surfacebrightness}
 \end{centering}
 \end{figure}

\begin{figure}[!ht]
 \begin{centering}
 \includegraphics[width=0.45\textwidth]{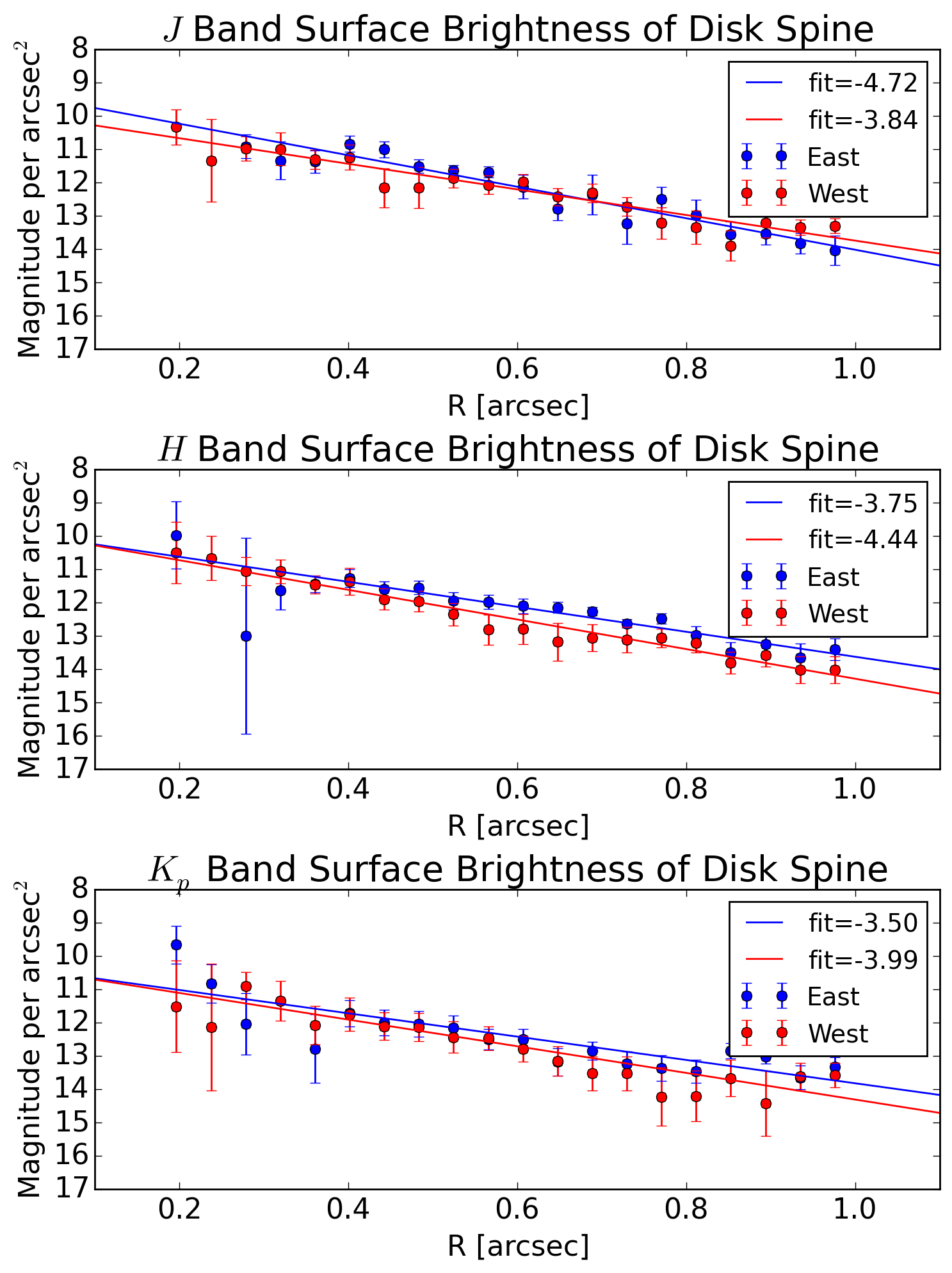}
 \caption
 {From top to bottom are plots of the $J$, $H$, and $K_p$ band surface brightnesses
 of the disk. The brightness asymmetry of the east and west sides of the disk are visible
 in these plots, albeit at differing separations and significance.}
 \label{fig:sb2}
 \end{centering}
\end{figure}

Figures~\ref{fig:surfacebrightness} and~\ref{fig:sb2} show the surface brightnesses/reflectance on the
east and west sides of the disk for the three color bands. The uncertainties decrease
significantly at radial separations of $\gtrsim$ 0\farcs{}25. These measurements
extend the surface brightness measurements inward from those calculated by~\cite{Thalmann2013}.  
The reflectance of the disk (surface brightness magnitudes - star's magnitudes) is slightly 
($\sim 1$ magnitude) blue at most radial separations.
Figure~\ref{fig:sb2} clearly shows the excess $H$ band brightness of the east 
side of the disk compared to that of the west side. This asymmetry
appears present at $J$ band at a smaller inner separation and is marginal but
plausible at $K_p$ band at a larger separation.
The disk's surface brightness radial profile can be well fit with a power law with an exponential
decay term of $-4.04 \pm 0.46$.

\section{Discussion}\label{sec:discussion}
Our improved signal to noise and inner working angle compared to those of previous work
enabled us to better constrain HIP 79977's
disk parameters. Our fitted parameters agreed with those derived 
by~\citet{Engler2017} within $1 \sigma$ except for the fiducial radius, which differs by $1.4 \sigma$ (this
takes into account the different distance they assumed). While our picture of the disk qualitatively agrees with much of that from the discovery paper \citep{Thalmann2013}, we exclude some of the parameter space for dust scattering that they find (e.g. $g$ = 0.4) and find a larger disk radius than they adopted in their paper ($r_0=40$ au).

\begin{figure}[!ht]
 \begin{centering}
 \includegraphics[width=0.4\textwidth]{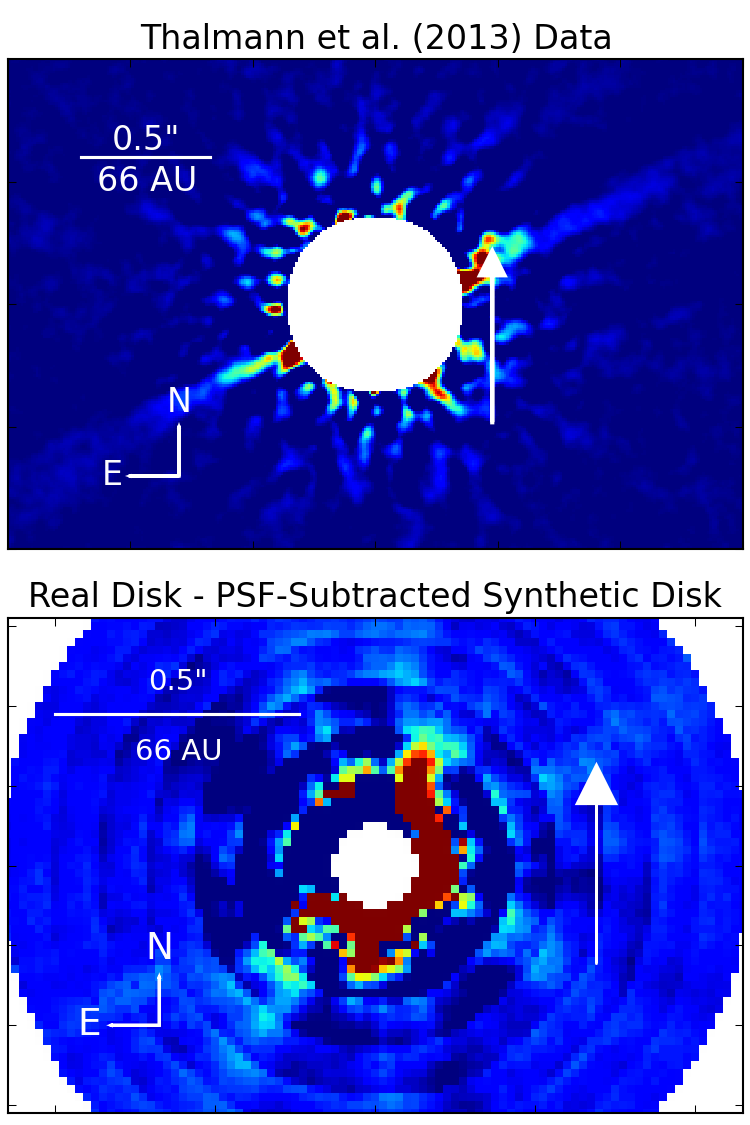}
 \caption
 {Top panel: reduction of the \citet{Thalmann2013}
 HIP 79977 data. An arrow points to the $4.6\sigma$ significance clump in their data.
 Bottom panel: our residuals after the forward-modeled best-fitting synthetic
 disk has been subtracted from the image. The same location is indicated with an arrow.}
 \label{fig:planet}
 \end{centering}
 \end{figure}

\citet{Thalmann2013} also note a candidate point source-like emission peak
located 0\farcs{}5 from the star, which appeared after subtracting their best-fit disk model.  They
posited that, if confirmed, this peak could be a localized clump of debris or
thermal emission from a $3-5 M_J$ planet\footnote{SCExAO
is a rapidly evolving platform that achieved a significant performance gain in the months after our
data were taken (O. Guyon, T. Currie, 2018 unpublished).   Thus, we defer discussion of limits on 
direct planet detections for a future HIP 79977 paper reporting new, substantially better data.}.
While our rereduction of the \citeauthor{Thalmann2013} data likewise show this emission, it does not
appear in the SCExAO/CHARIS data (Figure~\ref{fig:planet}) nor in the 2016 SCExAO/HiCIAO data.   
Given that both SCExAO data sets yield significantly deeper contrasts, we conclude that the emission
peak seen in AO188 data is likely residual speckle noise whose brightness highlights the stiff 
challenges in interpreting high-contrast imaging data where significant residual noise 
remains.
 
\begin{deluxetable*}{lcccccr}[!ht]
\tablecaption{Scattered Light Resolved Debris Disks Around 5-30 Myr old Stars \label{table:diskscompared}}
\tablecolumns{7}
\tablenum{2}
\tablewidth{0pt}
\tablehead{
\colhead{Star Name} & \colhead{Other} & \colhead{Age} & \colhead{$r_0$} & 
\colhead{H-G Parameter} & \colhead{Inclination} & \colhead{References} \\
\colhead{} & \colhead{Name} & \colhead{(Myr)} & \colhead{(au)} & 
\colhead{$g$} & \colhead{$i\; (\arcdeg)$} & \colhead{}
}
\startdata
 HD 146897 & HIP 79977 & 11 & 53 & 0.6 & 84.6& \citet{Thalmann2013}, this work \\\\
 GSC 0739-00759 & - & 23 & 70 & 0.50 & 83& \citet{Sissa2018}\\
 HD 15115 & HIP 11360 & $<$100 & 90 & 0.25 & 86.2& \citet{Kalas2007}, \citet{Mazoyer2014} \\
 %
 HD 36546 & HIP 26062 & 3-10 & 85 & 0.85 & 75 & \citet{Currie2017a}\\
 HD 39060 & $\beta$ Pic & 23 & 24--140 & 0.74 &85.2 & \citet{SmithTerrile1984}, \\
  & & & & & & \citet{MillarBlanchaer2015} \\
 HD 95086 & HIP 53524 & 17 &100--300& - & & \citet{Chauvin2018} \\
 %
 HD 106906 & HIP 59960 & 13 & 65 & 0.6 & 85.3& \citet{Lagrange2016}\\
 HD 109573 & HR 4796A & 10 & 77 & --\tablenotemark{a} & 76.5 & \citet{Schneider1999}, \citet{Milli2017}\\
 HD 110058 & HIP 61782 & 17 & 32 & - & $\sim$ 90? & \citet{Kasper2015}\\
 HD 111520 & HIP 62657 & 17 & 40-75 & - & 88?& \citet{Draper2016} \\
 HD 114082 & HIP 64184 & 16 & 26-31\tablenotemark{b} & 0.07-0.23\tablenotemark{b} & 82.3 & \citet{Wahhaj2016} \\
 HD 115600 & HIP 64995 & 15 & 48 & 0 &79.5& \citet{Currie2015a}\\
 HD 120326 & HIP 67497 & 16 &59, 130\tablenotemark{c} & 0.82, -\tablenotemark{c} & 80 & \citet{Bonnefoy2017} \\
 HD 129590 & HIP 72070 & 10-16 & 59 & 0.43 & 75& \citet{Matthews2017} \\
 HD 131835 & HIP 73145 & 15 & 90 & 0.15 & 75.1 & \citet{Hung2015}, \citet{Feldt2017} \\
 HD 181327 & HIP 95270 & 23 & 88 & 0.3\tablenotemark{d} &31.7 & \citet{Schneider2006}, \\
  & & & & & & \citet{Schneider2014} \\
 HD 197481 & AU Mic & 23 & 40-50 & $>0.7$\tablenotemark{e} & $\sim$ 90? & \citet{Kalas2004}, \citet{Graham2007} \\ 
 TWA 7 & CE Ant & 10 & 25 & 0.63 & 13& \citet{Choquet2016}, \citet{Olofsson2018} \\
 TWA 25 & V1249 Cen & 7-13 & 78 & 0.7 & 75 & \citet{Choquet2016} \\
\enddata
\tablenotetext{a}{Note that a Henyey-Greenstein scattering function fails to reproduce this disk's scattering phase function \citep{Milli2017}.   See Discussion.  }
\tablenotetext{b}{\citet{Wahhaj2016} reported values for three different data reductions, 
and we summarized their range of outcomes. Also,
instead of parameterizing the disk with $r_0$, inside and outside of which the disk drops off
in brightness, they assumed constant brightness between $r_{in}$ and $r_{in} + \Delta r$, with
falloff outside this range, and fit for both parameters. We reported their mean ring thickness 
$r_{in}+\frac{1}{2} \Delta r$.}
\tablenotetext{c}{\citet{Bonnefoy2017} detected two rings around HIP 67497 and modeled for both of them.}
\tablenotetext{d}{In their discovery paper,~\citet{Schneider2006} reported that HD 181327 had $r_0=86$
au and $g=0.3$. Later data modeled by \citet{Schneider2014} found $r_0=88$ au and surface brightness asymmetries that were not
well parameterized by a Henyey-Greenstein scattering function$g$ \citep[see also ][]{Stark2014}.}
\tablenotetext{e}{Au Mic has been extensively studied since~\cite{Graham2007}. However, publications
since then then have stopped fitting for $r_0$ and $g$ and have instead focused on characterization
of finer structures in the disk~\citep[e.g.][]{Boccaletti2018}.}
\tablecomments{
References are given for the first peer-reviewed publication of resolved optical/NIR imaging of the disk and the most recent paper that fitted for $r_0$ and $g$.
We report the age and best-fitting values of $g$ and $r_0$ from the second cited paper, unless there
has only been one publication, in which case we use its values.}
\end{deluxetable*}

Table \ref{table:diskscompared} casts the derived dust scattering properties and radius for HIP 79977's debris disk within the context of other scattered light
resolved debris disks around young (5--30 Myr old) stars that have been observed at near-infrared wavelengths.   Our 
derived value of the Henyey-Greenstein parameter ($g=0.6$) 
falls in the middle to upper end of the range observed for other debris disks resolved in scattered 
light around 5--30 Myr old stars.
The fiducial radius of the HIP 79977 disk is fairly typical of values measured for other
debris disks.   Taking both parameters together, the location and dust scattering properties of the HIP 79977 disk appear most similar to that for HD 106906 \citep{Lagrange2016}, GSC 07396-00759 \citep{Sissa2018}, and TWA 25 \citep{Choquet2016}.   In particular, HD 106906's disk is likewise best modeled (within the Henyey-Greenstein formalism) by strongly forward-scattering dust and exhibits a clear east-west brightness asymmetry, similar to what our data hint at for HIP 79977.

However, HIP 79977's derived dust scattering parameter need not imply that its dust is \textit{intrinsically} more forward-scattering than that of other young, resolved debris disks. 
Early studies employing a single Henyey-Greenstein scattering function implied neutral dust grains \citep[$g$ $\lesssim$ 0.16][] {Schneider2009,Thalmann2011}.   However, more recent analysis based on extreme-AO observations probing small scattering angles showed that the disk's scattering function is not
well-fit by a single Henyey-Greenstein parameterization but by a weighted combination of a strongly forward-scattering and strongly backward-scattering H-G component \citep{Milli2017}.  Further improvements to scattering phase functions may require departures from standard Mie theory, e.g.  Distribution of Hollow Spheres \citep[e.g.][]{Milli2017}.   

Furthermore, as shown in \citet{Hughes2018}, the derived H-G $g$ value strongly correlates with the range of probed scattering angles: the closer to the forward-scattering peak probed by the data, the higher the derived $g$ value.  Indeed, all of the ostensibly strongly forward-scattering disks listed in Table \ref{table:diskscompared} are highly inclined, where such small angles are accessible.   If there is little intrinsic difference in the scattering properties of young debris disks, then a single scattering phase function \citep[e.g.][]{Hong1985} should be able to reproduce the available data. On the other hand, higher quality data for other ostensibly neutral scattering disks like HD 115600 \citep{Currie2015a} should likewise reveal a forward-scattering component inconsistent with the Henyey-Greenstein formalism.


The disk flux in our images is scattered primarily by dust grains that are micron-sized and larger.
Grains much smaller than our observing wavelengths scatter light isotropically, whereas larger
grains preferentially forward scatter light~\citep{Hughes2018}. Therefore, if the disk
was dominated by grains with sizes smaller than a micron, we would not expect to have observed the 
forward scattering that we did.
On the other hand, grains smaller than the observing wavelength
scatter light in the Rayleigh regime and should produce blue colors, which is nominally more consistent with our results.
The quality of our data
limits our ability to make further inferences about the dust properties; the disk's
dust properties could be better constrained by resolved spectra with higher signal to noise than 
our observations or multi-band polarimetric analysis.

A possible brightness asymmetry appears in at least H band 
and seems plausible from the 2016 HiCIAO data
(Figure~\ref{fig:diskcomparison}c) and is broadly consistent with the ALMA dust continuum image probing much larger grains, which may show a slight asymmetry as well \citep[see Figure 1 in][]{LiemanSifry2016}.   However, it  will require confirmation with additional data sets of greater depth.
If confirmed, there are several plausible physical explanations for this emission asymmetry. 
An eccentric disk could the east side of the disk closer to us and appear brighter, although our forward-modeling suggests that the disk is consistent with having zero eccentricity thus far.
Alternatively, brightness asymmetries visible in a single band could identify compositional gradients across the disk \citep{Debes2008}; collisions of the debris
in the disk could produce lumpiness and anisotrophies of brightness, and these would fade away on the
dynamical timescale of the disk. 
While ~\cite{Engler2017}
did not identify this brightness asymmetry, their data were at optical wavelengths and in polarized intensity.

The surface brightness power law measured in Section~\ref{sec:colors} is consistent with the
disk model proposed by~\citet{Strubbe2006}. They suggest that at the fiducial radius $r=r_0$,
micron-size grains are produced by the collisions of parent bodies with circular orbits. Outward
of this radius lie grains large enough to remain gravitationally bound to the star but having orbits that
have become eccentric due to stellar winds and radiation pressure from the star. This model
produces a surface brightness profile that drops off beyond the fiducial radius as 
$r^{-\alpha}$, where $\alpha \approx 4-5$. This agrees with our measured value of $-4.1 \pm 0.4$.


Since the acquisition of the data presented in this paper, SCExAO has achieved significant performance improvements, reaching in excess of 90\% Strehl at 1.6 $\micron$ for bright stars \citep{Currie2018b}.   Thus, future, deeper SCExAO/CHARIS observations of HIP 79977 will enable a more robust characterization of the HIP 79977 disk's morphology and access the inner 0\farcs{}25 with higher signal to noise.   Multi-wavelength photometry  obtained from these data can identify color gradients in the disk possibly traceable to different dust properties \citep[e.g.][]{Debes2008}.   These photometric points, complementary $L_{p}$ imaging, and spatially-resolved spectra can provide crucial insights into how the morphology and composition of HIP 79977's debris disk compare to the Kuiper belt and other debris disks probing the epoch of icy planet formation \citep[e.g.][]{Currie2015a,Rodigas2015,Milli2017}.

\acknowledgments
We thank the anonymous referee for helpful suggestions that improved the quality of this work.   We also thank Laurent Pueyo for helpful conversations about KLIP forward-modeling. TC is supported by a NASA Senior Postdoctoral Fellowship.  MT is partly supported by the JSPS Grant-in-Aid
(15H02063). SG is supported from NSF award AST 1106391 and NASA Roses APRA award NNX 13AC14G.
The development of SCExAO was supported by the JSPS (Grant-in-Aid for Research \#23340051, \#26220704, \#23103002), the Astrobiology Center (ABC) of the National Institutes of Natural Sciences, Japan, the Mt Cuba Foundation and the directors contingency fund at Subaru Telescope.  
CHARIS was built at Princeton University in collaboration with the
National Astronomical Observatory of Japan under a Grant-in-Aid for
Scientific Research on Innovative Areas from MEXT of the Japanese
government (\#23103002).
We wish to emphasize the pivotal cultural role and reverence that the summit of Maunakea has always had within the indigenous Hawaiian community.  We are most fortunate to have the privilege to conduct scientific observations from this mountain. 

%
\vspace{5mm}
\facilities{Subaru Telescope (SCExAO, CHARIS)}
\software{CHARIS Data Reduction Pipeline
          }

\end{document}